\documentclass[final,leqno,onefignum,onetabnum]{scrartcl}

\usepackage{amsmath}
\usepackage{amsfonts}
\usepackage{amssymb}

\usepackage{graphicx}
\usepackage{caption}
\usepackage[position=b]{subcaption}

\renewcommand{\vec}{\mathbf}
\renewcommand{\eqref}[1]{Eq.~(\ref{#1})}
\newcommand{\figref}[1]{Fig.~\ref{#1}}

\usepackage{relsize}
\usepackage{xspace}
\protected\def\plusplus{{\nolinebreak[4]\hspace{-.05em}\raisebox{.4ex}{\relsize{-3}\bf ++}}\xspace}
\newcommand{\CXX}{{\rm C}\plusplus}

\newcommand{\stellar}{{STE\raisebox{.05em}{\hspace{.1em}$|$\hspace{0.1em}$|$\hspace{.02em}}AR}\xspace}

\title{Optimizing Large-Scale ODE Simulations}

\author{Mario Mulansky\thanks{Institute for Complex Systems, CNR, Sesto Fiorentino, Italy 
}
}

\begin{document}

\maketitle

\begin{abstract}
We present a strategy to speed up Runge-Kutta-based ODE simulations of large systems with nearest-neighbor coupling.
We identify the cache/memory bandwidth as the crucial performance bottleneck.
To reduce the required bandwidth, we introduce a granularity in the simulation and identify the optimal cluster size in a performance study.
This leads to a considerable performance increase and transforms the algorithm from bandwidth bound to CPU bound.
By additionally employing SIMD instructions we are able to boost the efficiency even further.
In the end, a total performance increase of up to a \emph{factor three} is observed when using cache optimization and SIMD instructions compared to a standard implementation.
All simulation codes are written in \CXX and made publicly available.
By using the modern \CXX libraries Boost.odeint and Boost.SIMD, these optimizations can be implemented with minimal programming effort.
\end{abstract}

%

\section{Introduction}
The 
numerical integration of ordinary differential equations (ODEs) is a frequent and highly important task in all scientific areas.
The efficient implementation of ODE simulations therefore plays a crucial role in many scientific applications as for example seen from the numerous scientific libraries devoted to solving ODEs, e.g.~\cite{shampine1997matlab,hindmarsh2005sundials,press2007numerical,Ahnert_Mulansky_odeint:11}.
In recent years, the available computational power has continued to grow tremendously, but the utilization of this power introduces new challenges, especially in the field of numerical simulations where optimal performance is often crucial.
The most prominent example is GPGPU computing that has become very popular in the past years, and for which the applicability of ODE simulations was shown recently~\cite{murray2012gpu,demidov2013programming,dindar2013swarm,ahnert2014solving}.
Besides the highly parallelized GPUs containing hundreds of cores, also normal CPUs now consist of many cores, and modern workstations include several CPUs.
Hence it seems that nowadays performance gains are best achieved by parallelization of the algorithms and thus utilizing all available cores.
But this is short-sighted.
To reach optimal runtime one always should start tuning the single-thread performance before considering parallelization and multi-threading.
This is a rather obvious fact, as an optimized single-thread performance is also beneficial for possible later parallelization of the algorithm.
However, current processors are highly complex and although they can provide enormous computational power, it has become increasingly difficult to utilize their full performance.

An often overlooked problem in numerical computing is memory bandwidth limitation.
In many cases programs are not bound by the available computational power (CPU throughput), but rather by the fact that the CPU has to wait for the required data to arrive from the main memory.
Indeed, memory access typically has a latency of the order of a hundred CPU cycles, which can introduce enormous performance losses.
Therefore, CPUs are equipped with several layers of caches: L1, L2 and in modern processors also L3 cache.
Those caches can pre-load the required data for the CPU with the purpose of reducing latencies due to main memory access.
However, many algorithms are implemented in a way that makes efficient cache usage hard or even impossible.
In such situations, adjusting the algorithm to reduce the required cache and/or memory bandwidth can greatly improve the performance.
The importance of cache tuning is known since a long time~\cite{kagi1996memory,lebeck1994cache} and has become increasingly relevant as the computational power continued to grow faster than the memory bandwidth~\cite{ding2000memory}.
Consequently, many modern numerical libraries address this problem and include optimization techniques for enhancing cache performance~\cite{kowarschik2003overview,gunther2006cache}.
In the following, we will see that to reduce memory bandwidth, it can be even beneficial to repeat parts of the computation instead of accessing previous results.

But even if the algorithm is cache friendly and limited by the CPU throughput, there is room for improvement.
Modern processors include SIMD extensions (Single Instruction Multiple Data) that allow the parallel computation of two or four double-precision operations in a single clock-cycle on a single core~\cite{patterson2013computer}.
Although compilers are capable of making use of SIMD instructions automatically, a process typically called auto-vectorization, the specific use of SIMD instructions by the programmer can give further significant performance gains as will be also demonstrated here.

In the following sections, we present a new technique to increase the performance of large ODE simulations of \emph{one-dimensional chains with nearest-neighbor interactions}.
The speed up is based on reducing the required cache bandwidth by introducing a granularity in the algorithm.
By adjusting the granularity we will be able to optimize the cache usage and thus increase the performance significantly.
We will then make use of SIMD instructions to further tune the efficiency and reach a speed up of up to a factor of three compared to the standard implementation (cf.\ Fig.~\ref{fig:speedup}).

\section{Optimizing Simulation Performance}
\enlargethispage{1em}

We address the quite general situation of a high-dimensional ODE with nearest-neighbor interactions.
The numerical approximation of the solution will be computed with the explicit Runge-Kutta-4 method~\cite{press2007numerical}.
This is the standard, general purpose routine and widely used due to its robustness and simple implementation.
However, the techniques presented below can be applied also to any other explicit Runge-Kutta method.

\subsection{Model with Nearest-Neighbor Coupling}
Let us introduce a chain with nearest neighbor interactions as follows: the state of the $i$-th element of the chain is represented by $\vec r_i(t)$, in general a $D$-dimensional vector.
The chain has length $N$, i.e.\ $i=1\dots N$, hence in total the system is described by $N\cdot D$ scalar values.
The independent variable $t$ is usually representing the time in case of dynamical systems.
Most generally, an ODE with nearest-neighbor interactions can be written as:
\begin{equation}
 \dot {\vec r}_i = \vec f_i(\vec r_i, t) + \vec g_{i} (\vec r_i, \vec r_{i-1}, \vec r_{i+1}, t),
\end{equation}
where $\dot {\vec r}$ denotes the derivative with respect to $t$ and we have omitted the explicit time-dependence $\vec r(t)$.
The function $\vec f_i$ represents the local term and $\vec g_{i}$, the nearest-neighbor coupling.
Note that in this setup both the local and the coupling term can be different for each site.
In many cases, however, one faces homogeneous situations where $\vec f$ and $\vec g$ are the same for all elements:
\begin{equation} \label{eqn:hom_nn_ode}
 \dot {\vec r}_i = \vec f(\vec r_i, t) + \vec g (\vec r_i, \vec r_{i-1}, \vec r_{i+1}, t).
\end{equation}
Here, we will also assume such a homogeneous situation.
Examples for such systems include nonlinear oscillator chains~\cite{Mulansky_Pikovsky_13,bordyugov2010self}, nonlinear Klein-Gordon models~\cite{flach2009universal}, the discrete nonlinear Schr\"odinger model~\cite{mulansky2013simulating} and classical spin chains~\cite{oganesyan2009energy}.
Furthermore, sometimes also discretized partial differential equations are solved by explicit schemes, e.g.~\cite{Wang_Li_Song_08}.
Nevertheless, the cache optimization technique presented below can be generalized to heterogenic cases, but will not work with implicit routines.

\subsection{Runge-Kutta Schemes}

\begin{figure}[t]
  \centering
    \includegraphics[width=0.7\textwidth]{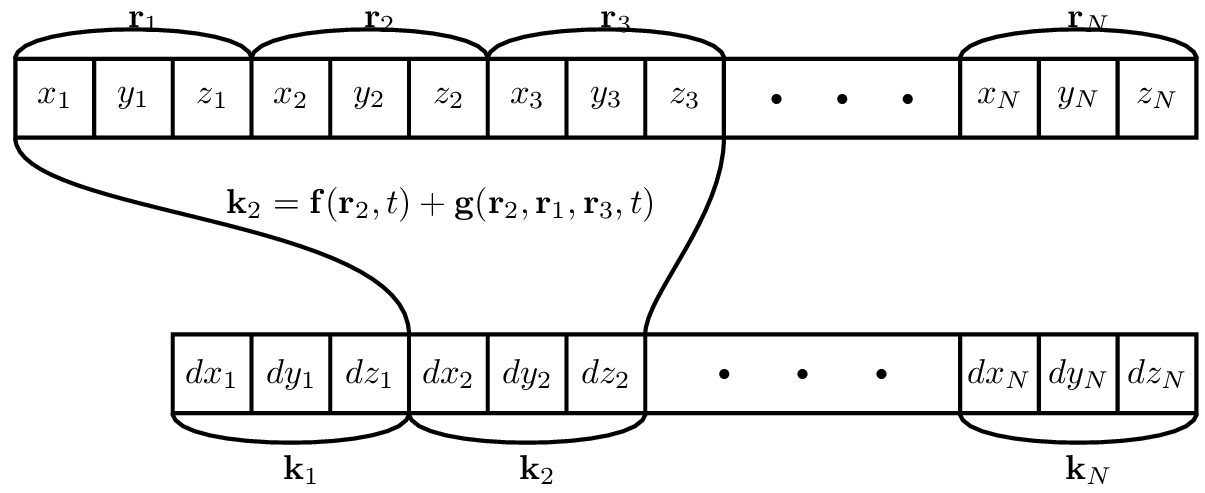} \hfill
  \caption{Memory layout and data dependency of the ODE function evaluation}
   \label{fig:memory}
\end{figure}

To simulate such a system with nearest-neighbor coupling, we first have to choose a memory layout for representing the state $\{\vec r_i\}$.
To avoid unnecessary cache misses, the best strategy is to place the data required at the same time in the algorithm also close in memory.
For a system with nearest-neighbor coupling this is quite simple: the local function $\vec f$ needs all elements of a single state $\vec r_i$, while the coupling $\vec g$ additionally requires the left and right neighbors.
So the best memory layout is very naturally to put all elements of $\{\vec r_i\}$ consecutively  one after another.
This is shown in \figref{fig:memory} for an example with $D=3$, i.e.\ $\vec r_i = (x_i, y_i, z_i)$.
There, it is also illustrated how all data required to evaluate the rhs of the ODE for one element ${\vec k_i}$ (see \eqref{eqn:rhs_k} below) is found in a continuous memory block.
Only at the boundary there is potentially an access to the other end of the chain if periodic boundary conditions are employed.
However, in the case of large systems, which are considered here, the boundary becomes negligible for the overall performance.

\begin{figure}[t]
   \centering
   \includegraphics[width=0.5\textwidth]{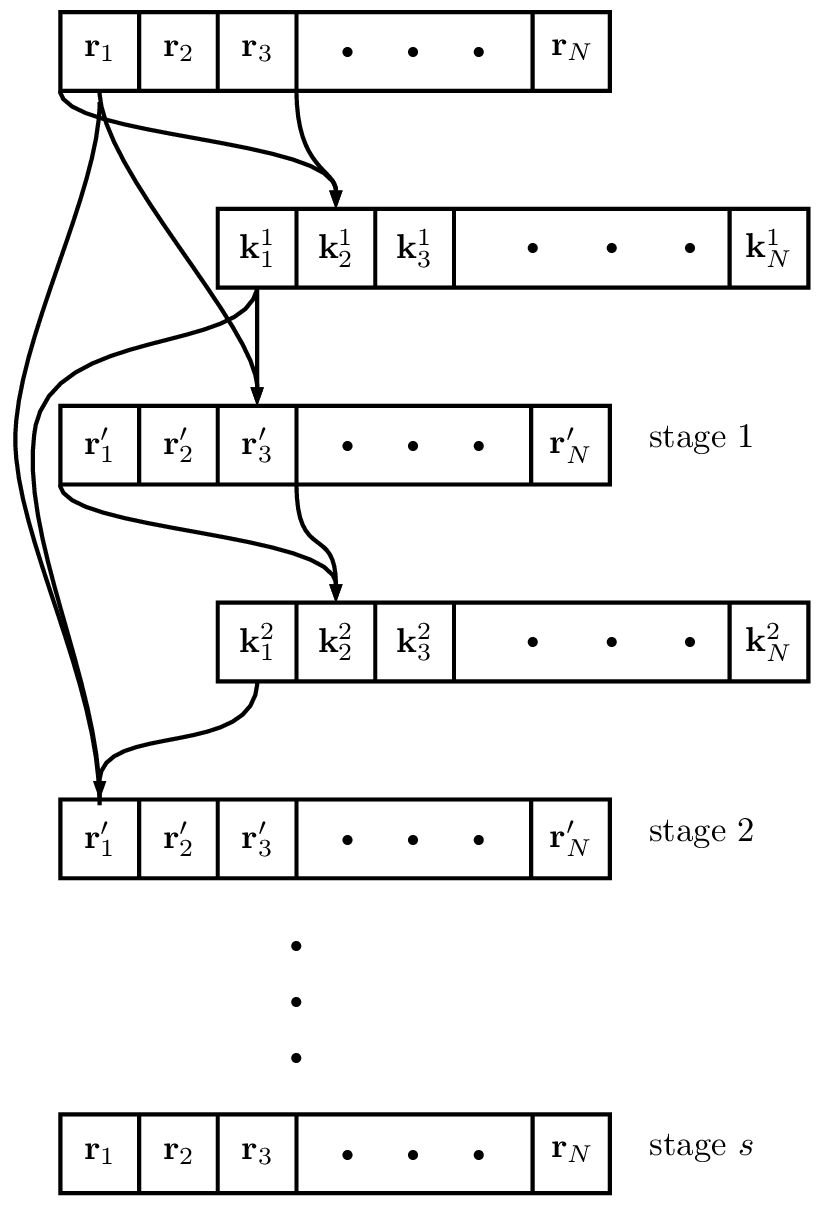}
   \caption{Schematic flow of a Runge-Kutta scheme.\label{fig:rk4}}
\end{figure}

The implementation of an explicit Runge-Kutta scheme is rather simple.
Starting from some given initial condition $\{\vec r_i\}$ for the time $t$ we will find an approximate solution $\{\bar{\vec r}_i\}$ at $t+\Delta t$ for some small step size $\Delta t$.
The calculation consists of several stages, where at each stage $j$ first an evaluation of the rhs of the ODE is performed:
\begin{equation} \label{eqn:rhs_k}
 \vec k^j_i = \vec f(\vec r_i,t)+\vec g(\vec r_i, \vec r_{i-1}, \vec r_{i+1}, t+c_j\Delta t).
\end{equation} 
Then, a new intermediate approximation is computed from the current, and possibly previous, rhs evaluations:
\begin{equation} \label{eqn:rk_intermediate}
 \vec r'_i = \vec r_i + \sum_{n=1}^j a_{j,n} \Delta t\, \vec k^n_i.
\end{equation}
This intermediate approximation is then used in the next stage to compute another rhs evalution $\{\vec k^{j+1}_i\}$ and so on.
\figref{fig:rk4} schematically shows the flow of the algorithm.
After $s$ stages the final result is obtained using the previously computed evaluations of the rhs:
\begin{equation} \label{eqn:rk_final}
 \bar{\vec r}_i = \vec r_i + \sum_{n=1}^s b_{n} \Delta t\, \vec k^n_i.
\end{equation} 
The number of stages $s$ as well as the parameters $a_{j,n}$ and ${c_j}$ are properties of the Runge-Kutta scheme.
The prominent Runge-Kutta-4 scheme for example has $s=4$ stages, i.e.\ four evaluations of the rhs function in \eqref{eqn:rhs_k} per time step.


Implementing this algorithm is rather straight-forward and several libraries exist offering such functionality~\cite{shampine1997matlab,hindmarsh2005sundials,press2007numerical,Ahnert_Mulansky_odeint:11}.
However, all those implementations potentially suffer from severe performance loss when dealing with large systems due to memory bandwidth limitations.
For example, evaluating the rhs of the ODE \eqref{eqn:rhs_k} using double precision (8 Bytes per value) has a memory requirement of $2\cdot N\cdot D\cdot 8\, \text{Byte}$.
For optimal performance, this data should be available in the L1 or L2 cache of the CPU, as those caches can be accessed by the CPU with low latency and extremely high bandwidth.
However, those caches have only limited size, typically less than a few hundred Kilobytes.
For large systems, i.e.\ $N\cdot D\gtrsim 10^5$, the memory requirement of the state $\{\vec r_i\}$ and the rhs $\{\vec k^j_i\}$ outrun the available L1/L2 cache and the CPU has to employ the L3 cache, which has a larger latency and much lower bandwidth.
For very large systems, $N\cdot D\gtrsim 10^7$, eventually also the main memory has to be used resulting in even more significant bandwidth limitations.
That means, for large systems \emph{any} iteration over $\{\vec r_i\}$ or $\{\vec k^j_i\}$ is slowed down by cache latency and/or memory bandwidth.

The Runge-Kutta scheme involves many such iterations.
For each stage, the rhs evaluation \eqref{eqn:rhs_k} as well as the computation of the intermediate approximation \eqref{eqn:rk_intermediate} involves an iteration over all $N$ elements, hence a total of $2s$ such iterations.
This can lead to a severe performance loss as shown in \figref{fig:perf_ariel}.
There, we use ${N=2^{20}\approx10^6}$ and $D=3$ which means one state $\{\vec r_i\}$ occupies 24 Megabytes, much more than the available L1/L2 cache size.
And indeed, we observe a significant activity for the L3 cache (bottom panel in \figref{fig:perf_ariel}) for the standard implementation, which results in a sub-optimal performance.
More details on the performance measurements follow below.
First, we will provide an approach to circumvent this problem by introducing granularity to the algorithm.

\subsection{Introducing Granularity}
As described above, every iteration across all $N$ elements can lead to performance loss if the state becomes too large.
The solution is rather simple: divide the state into clusters such that each cluster fits into L1/L2 and do a whole Runge-Kutta step on each cluster in turn.
Instead of $2s$ iterations for each Runge-Kutta step this would involve only a single iteration and therefore reduce the required L3/memory bandwidth by a factor of $2s$.
For an uncoupled system, i.e.\ $\vec g=0$, this is trivial as all systems are independent from each other and thus can be treated separately.
In the presence of coupling, however, the situation is more complicated.
Suppose we divide the state into $C$ clusters, each having $G$ elements, i.e.\ $N=C\cdot G$.
The cluster size $G$ is also called \emph{granularity}.
Now we want to perform one Runge-Kutta step for a single cluster that starts at index $g$ and ends at $\bar g-1$: $(\vec r_g,\dots, \vec r_{\bar g-1})$, where $\bar g=g+G+1$ is the start index of the next cluster.
Hence, we have to compute the function values $(\vec k^1_g,\dots, \vec k^1_{\bar g-1})$ in \eqref{eqn:rhs_k}.
But in general (arbitrary coupling), for this computation we need all values $\{\vec r_i\}$, not only those from the cluster $g$ due to the coupling terms.

\begin{figure}[t]
   \centering
   \includegraphics[width=0.75\textwidth]{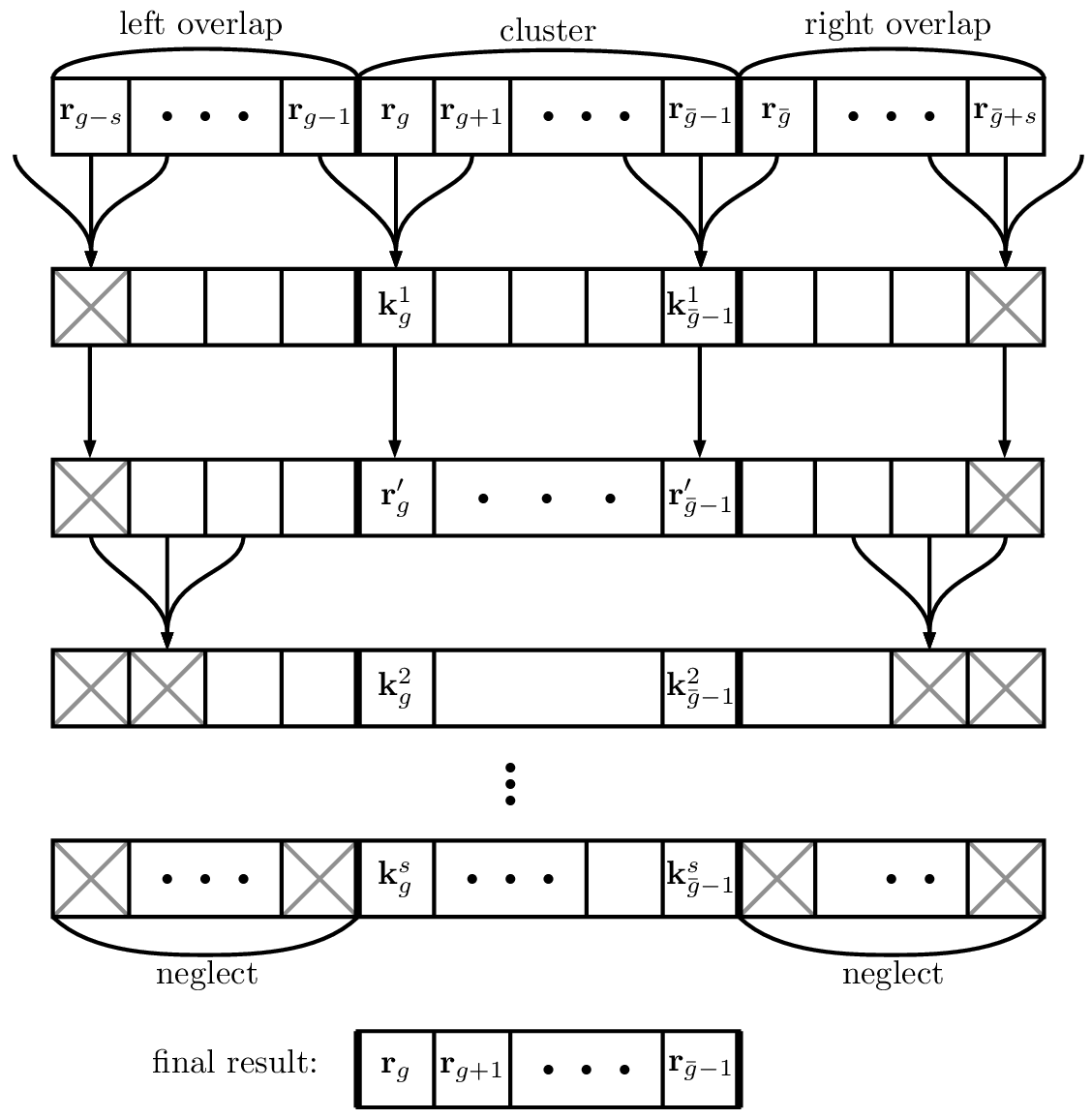}
   \caption{Error propagation in the Runge-Kutta scheme.\label{fig:rk4_cluster}}
\end{figure}

For nearest-neighbor coupling, the situation is much better.
Here, we are only missing the values $\vec r_{g-1}$ and $\vec r_{\bar g}$ to compute the values at the left and right boundary of the cluster: $\vec k^1_g$ and $\vec k^1_{\bar g-1}$.
So we simply append those values to the left and right end of the cluster.
Although we are only interested in the result $(\vec{\bar r}_g,\dots,\vec{\bar r}_{\bar g-1})$, we perform the Runge-Kutta step on the extended cluster $(\vec{\bar r}_{g-1},\dots,\vec{\bar r}_{\bar g})$.
However, this only fixes the first stage of the Runge-Kutta step.
That means we obtain a correct intermediate approximation for $(\vec r'_g,\dots, \vec r'_{\bar g-1})$, while the boundary values of the extended cluster $\vec r'_{g-1}$ and $\vec r'_{\bar g}$ are now wrong as for those again the required coupling terms are not available.
But for the next stage we need exactly those boundary values.
Therefore, we extend the cluster even further and add two neighbors on the left and right, which then gives correct results for two stages of the Runge-Kutta scheme.
For each stage of the algorithm the error introduced by the missing coupling term at the edges advances one index into the cluster, as sketched in \figref{fig:rk4_cluster}.
So finally, for a Runge-Kutta scheme with $s$ stages we have to extend the original cluster by $s$ left and right neighbors to ensure the correct computation of the cluster, as depicted in \figref{fig:rk4_cluster}.
In the end, we have the correct result for the cluster $(\vec r'_{g},\dots, \vec r'_{\bar g-1})$, while the extra values on the left and right are wrong and will be neglected.
Those values belong to the left (right) neighboring clusters and their correct computation will be performed typically before (after) the current cluster.
Consequently, we add an overlap computation of $2s$ values per cluster, which means an increase of the required computations by a factor of $1+2s/G$.
But we can hope that this increase of computations is out-weighted by the reduced cache/memory access of the algorithm.

And indeed, for an optimal choice of the granularity $G$, the cache optimization described above leads to a performance increase of roughly a factor of two, as shown examplarily in \figref{fig:perf_ariel} (``Cache optimized'' vs ``Standard'').
Some estimate for the optimal granularity can be deduced from the L1/L2 cache size.
However, there is a trade-off between cache usage and the amount of overlap computation and thus we employ performance study where different granularity values are compared, as shown in \figref{fig:granularity_ariel}.
Again, details on the performance results will follow below.

\subsection{SIMD Instructions}

As seen in \figref{fig:perf_ariel}, we can reach about 4200 Megaflops/s, that is 1.2 Flops/cycle on a 3.8 GHz Intel Xeon processor with the cache optimized implementation.
This is quite impressive, as typically the floating point unit of the CPU saturates at 1 Flop/cycle.
Hence, we were able to turn the memory bandwidth bound implementation into a CPU bound version by introducing granularity.
The extra performance comes from the SIMD (Single Instruction Multiple Data) units in the CPU, that can perform up to four floating point operations per cycle.
Modern compilers can utilize those extra SIMD registers automatically (auto-vectorization).
However, by explicitly using SIMD registers in the program one can hope to reach an even higher performance than by relying on auto-vectorization.

\begin{figure}[t]
  \centering
    \includegraphics[width=0.75\textwidth]{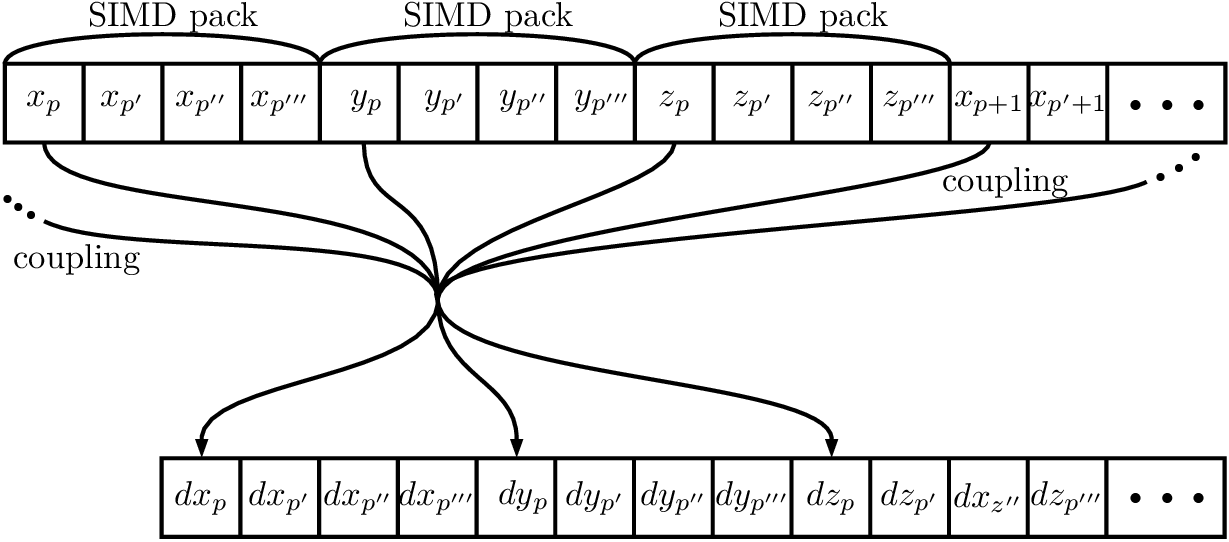} \hfill
  \caption{Memory layout of a chain of 3D systems organized in SIMD packs. Each pack contains $P=4$ values of the same type ($x$, $y$ or $z$ coordinate of a single system) taken from distant points in the chain (see text). Note how the evaluation of the rhs function for one value in a SIMD pack does not depend on other values of that SIMD pack.
  }
   \label{fig:memory_simd}
\end{figure}

For that we first have to adjust the memory layout of the implementation.
To make optimal use of the SIMD registers, we will organize the data in packs of size $P=4$.\footnote{We use a pack size of four as the target processor supports the AVX extensions with 256 Bit registers capable of holding four values (double precision). Older CPUs might only support SSE with 128 Bit registers, where a pack size of two should be used.}
To allow for optimal utilization of the SIMD units, we have to ensure that for all elements of an SIMD pack the same operations are performed, and that the operation of one entry in the pack does not depend on any other entry of the same pack.
Only then can the code be fully vectorized with significant performance gains.
Again assume we have a chain of three-dimensional systems ($D=3$), i.e.\ $\vec r_i=(x_i,y_i,z_i)$, and a chain length of $N$: $i=1,\dots,N$.
To make sure the same operations are performed on all entries of a pack we should only put either $x$, $y$, or $z$ values in a pack, e.g.\ $\{x_1,x_2,x_3,x_4\}$, $\{y_1,\dots,y_4\}$ and so on.
But with this organization, the computation of the rhs of the ODE in \eqref{eqn:rhs_k} would introduce coupling between the entries of the pack.
To avoid that, we divide the chain in $P=4$ pieces of length $N'=N/P$ and represent each piece by one entry of the SIMD packs.
So let $p=1$, $p'=N'+1$, $p''=2N'+1$ and $p'''=3N'+1$ be the starting indices of the four parts of the chain.
Then the first SIMD pack will contain $\{x_p, x_{p'}, x_{p''}, x_{p'''}\}$, the second one $\{y_p, y_{p'}, y_{p''}, y_{p'''}\}$ and so on.
This memory layout is sketched in \figref{fig:memory_simd}.
Compared to the original memory layout in \figref{fig:memory}, we replace the single values, e.g.\ $x_1$, by four values from distant points of the chain ($x_{p,p',p'',p'''}$).

Using such a memory layout, it is now possible to compute four values at once by virtue of SIMD.
In principle this could give a performance gain of a factor four in the best case.
However, note that this relies on the fact that the dynamical equations $\vec f +\vec g$ are indeed independent of the chain index, as the above usage of SIMD registers requires that the same operations are performed for each entry of the SIMD pack and hence in the above description also for each element in the chain.
Note also that at the edges of the four parts of the chain there is coupling between different entries of the SIMD pack.
For example the second chain is stored in the second SIMD entries, but the left neighbor of its first element is the last element of the first chain and hence stored in the first entries of the SIMD packs.
So the edges need special treatment and their calculation can not be carried out using the SIMD vectorization.
But again for long chains this has negligible performance impact.
Finally, the cache optimization described above can be similarly applied to this memory layout as well.
It works exactly in the same way, only that one now deals with SIMD packs instead of scalar values, but in return one has a reduced chain length of $N'=N/P$.

\figref{fig:perf_ariel} shows the performance comparison of the standard implementation, the optimized cache version and SIMD version.
As seen there, using SIMD instructions gives another performance boost of about 50\% compared to the cache optimized version, which means a total speed up of a factor of three compared to the standard implementation in this simulation.
Details of these performance tests are described below.

\section{Performance Study}
In the following, we will compare the performance of the different approaches presented above.
The examplary system is a chain of Roessler oscillators~\cite{Roessler1976397} with nearest-neighbor coupling.

\subsection{Coupled Roessler Chain}

A single Roessler oscillator has a three-dimensional (${D=3}$) state ${\vec r_i = (x_i,y_i,z_i)}$ following the dynamics:
\begin{equation}
\dot{\vec r}_i = \vec f(\vec r_i,t) = \left(
 \begin{array}{c}
  -y_i - z_i\\
  x_i + ay_i\\
  b + z_i(x_i-c)
 \end{array}
 \right)
\end{equation}
The parameters $a,b,c$ are constant and fixed to some typical values: $a=0.2$, $b=1$, $c=9$.
For those parameter values, the dynamics of the Roessler system is generally chaotic due to the existence of a strange attractor~\cite{Ott_chaos:2002}.
We add dispersive nearest-neighbor coupling in the first coordinate by defining:
\begin{equation}
 \vec g(\vec r_i, \vec r_{i-1}, \vec r_{i+1}) = \left(
 \begin{array}{c}
  x_{i-1} - 2x_i + x_{i+1}\\
  0 \\
  0
 \end{array}
 \right).
\end{equation}
At the boundaries, we set $\vec r_0 = \vec r_N$ and $\vec r_{N+1} = r_1$, i.e.\ periodic boundary conditions.
Substituting $\vec f$ and $\vec g$ in \eqref{eqn:hom_nn_ode} defines our chain of coupled Roessler oscillators that will be used for performance investigations below.


\subsection{Performance Results}

We simulate a chain of $N=2^{20}\approx 10^6$ coupled Roessler oscillators.
As mentioned above, we use the well-known and popular Runge-Kutta-4 algorithm with $s=4$ stages for the numerical time evolution.
We start from random initial conditions and perform $T=50$ steps with a step-size of ${\Delta t=0.01}$.
We measure the runtime $t_\text{Wall}$ needed to perform those $T$ steps and then quantify the performance in terms of Runge-Kutta steps per second:
\begin{equation}
 \text{Perf} = T/t_\text{Wall}.
\end{equation}
This is repeated for $M=10$ times and we use the minimal runtime (maximal performance) of those 10 trials.
Additionally, we measure the CPU throughput in terms of Flops/s (FLoating point OPerations per second) as well as the cache transfer bandwidth using the \texttt{likwid} framwork~\cite{treibig2010likwid}, which is based on the processor's performance counters.

\subsubsection{Results on Intel Xeon}
\begin{figure}[t]
  \subcaptionbox{Performance comparison.\label{fig:perf_ariel}}[0.48\textwidth]{
    \includegraphics[width=0.48\textwidth]{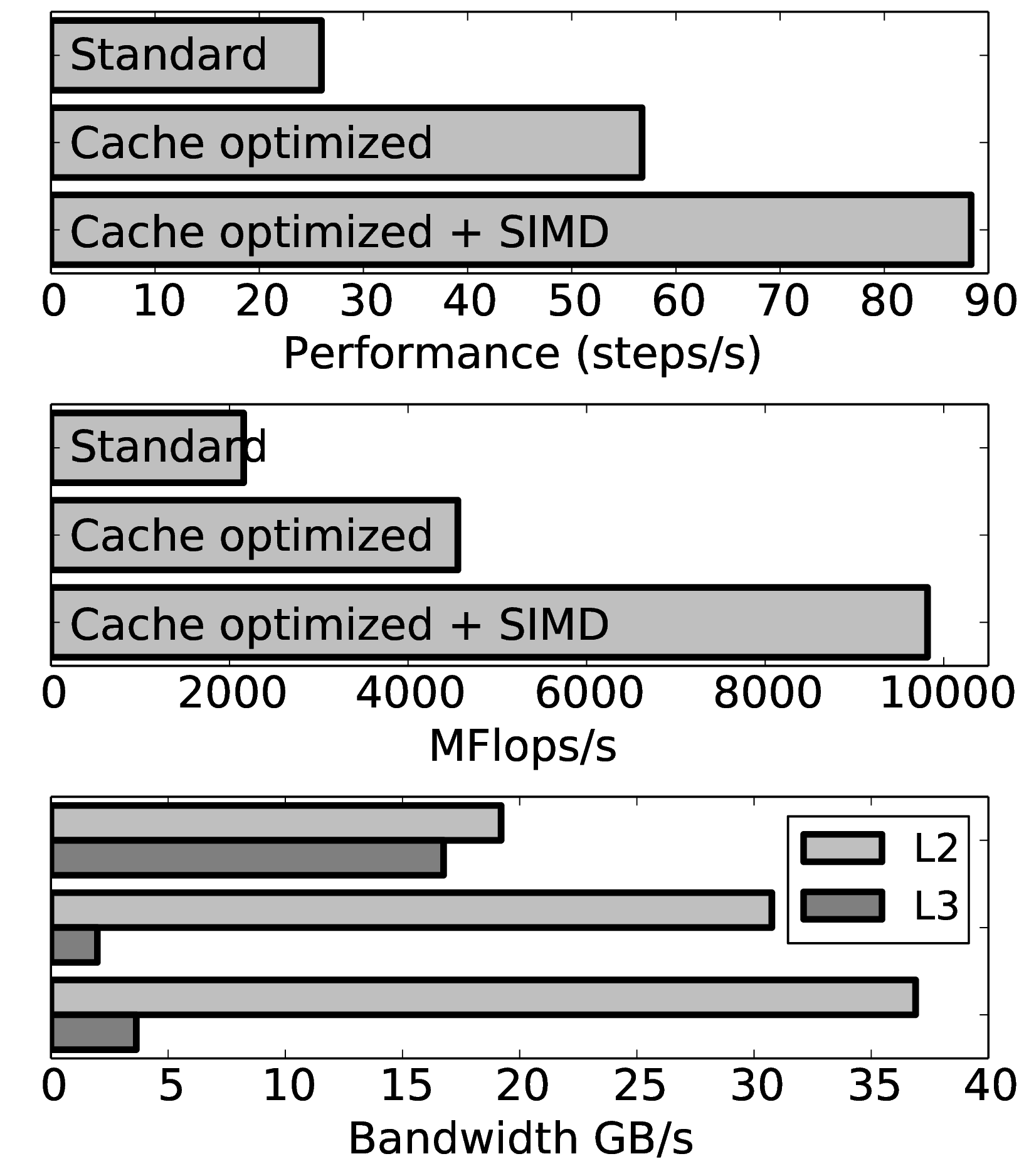}}\hfill  
  \subcaptionbox{Granularity dependence.\label{fig:granularity_ariel}}[0.48\textwidth]{
   \centering
   \includegraphics[width=0.48\textwidth]{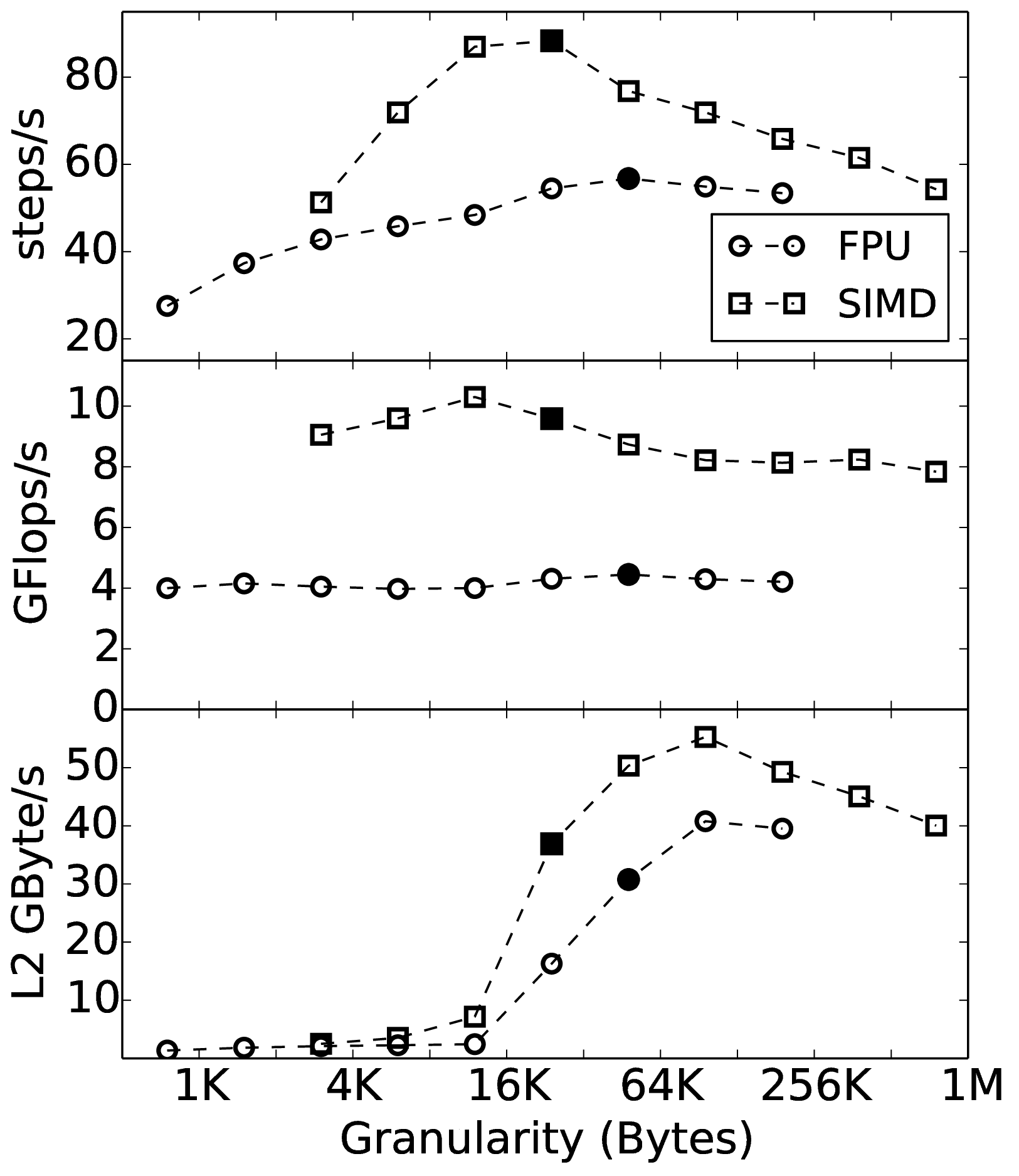}}
  \caption{Performance of a simulation of ${N=2^{20}\approx10^6}$ (24~Megabytes) coupled Roessler systems on an Intel Xeon 3.8~GHz. Left panel shows the performance of the three different versions: Standard implementation, cache optimized version and cache optimized + SIMD (both at optimal granularity). In the right panel the granularity dependence of the cache optimized version without (circles) and with (squares) additional SIMD usage is shown. The filled symbols represent the optimal value (highest performance).\label{fig:perf_ariel_both}}
\end{figure}
In a first run, the simulations are performed on an Intel Xeon E5 processor with 3.8 GHz (single core, turbo mode).
The simulations are implemented in \CXX and compiled with the Intel Compiler version 15.0.
The Intel Xeon processor has 32KB/256KB/20MB of L1/L2/L3 cache respectively.
The representation of one state $\{\vec r_i\}$ requires ${N\cdot D \cdot 8\, \text{Byte}} = 24\, \text{Megabytes}$, clearly above the L1/L2 cache capacity, but also bigger than the available L3 cache.
Therefore, we expect the standard implementation, where $2s=8$ iteration over the state are performed in each Runge-Kutta step, to suffer severely from bandwidth limitations.
This is seen in \figref{fig:perf_ariel}, where the top bar in all three panels represents this standard implementation.
We find a CPU throughput of roughly 2000 MFlops/s, which is significantly below the 3800 MFlops/s the FPU unit of this CPU is capable of.
Accordingly, the L3 cache bandwidth of about 17 GByte/s clearly indicates significant L3 usage.
Hence, the performance remains at below 30 Runge-Kutta steps per second.

Next, we analyze the performance of the cache optimized implementations introducing granularity and overlap computations as described above.
Therefore, in a first step we identify the optimal granularity $G$ by means of a performance study for different values~$G$.
The result is shown in \figref{fig:granularity_ariel} (circles), where we plot the performance (top panel), the CPU throughput (center panel) and the L2 cache bandwidth (bottom panel) for increasing granularity.
The granularity is measured in Bytes via ${D\cdot G\cdot 8\,\text{Bytes}}$ to allow for comparison with the cache size.
The analyzed range for $G$ spawns from values significantly smaller than the L1 cache size ($G\approx 1\,\text{KB}$) to about the L2 cache size ($G\approx1\,\text{MByte}$).
We find that the CPU throughput (center panel) is rather independent of the granularity and stays at about 4 GFlops/s.
The performance (top panel) however increases with increasing granularity because due to the larger sizes of the clusters (larger granularity), less overlap computation are performed and less Flops/s are wasted for neglected results.
The L2 bandwidth starts to increase at a granularity size of about 16~KByte, which is consistent with the L1 cache size of 32~KByte as for example a rhs computation needs to store the state as well as the result.
However, the L2 cache access does not have a severe influence on the performance as it is typically fast enough to supply the FPU with the required data.
Nevertheless, we identify the optimal granularity at 48~KBytes, which corresponds to a length of each cluster of $G=512$ elements.
\figref{fig:perf_ariel} then shows the performance results for this optimal value of the granularity (``Cache optimized'').
And indeed, we find a performance improvement as well as a CPU throughput improvement of a factor of two, while the L3 bandwidth drops by a factor of eight, as expected as now only one iteration over all clusters needs to be performed instead of $2s=8$ before.

Finally, we additionally employ the SIMD instructions as explained above.
Again we identify the optimal granularity in a similar fashion as before.
The results are shown as squares in \figref{fig:granularity_ariel}.
With SIMD instructions, the performance is more sensitive to the granularity than before and we find an optimal value of 24~KByte, which corresponds to an optimal cluster length of $G=64$ elements\footnote{Remember that each element of the chain in the SIMD case consists of a pack of four values.}.
Note, that with SIMD instructions, the L2 access does lead to a drop in the CPU throughput, as now potentially four times as much data is required per CPU cycle.
However, the optimal performance (top panel in \figref{fig:granularity_ariel}) does not correspond to the best CPU throughput (second panel) due to the overlap computations that contribute to the throughput but not to the real performance.
\enlargethispage{2em}

\figref{fig:perf_ariel} shows the performance, CPU throughput and cache bandwidths for the SIMD version at optimal granularity (``Cache optimized + SIMD'').
We observe a speed up of about 50\% over the plain cache optimized version and a factor three over the standard implementation.
The gain in CPU throughput is significantly higher, which can be partially explained by the fact that in this case more overlap computations have to be performed ($12.5\%$ at $G=64$ vs $\sim 1\%$ at $G=512$).
However, note that the performance counters used to measure the CPU throughput are not completely reliable.
Hence, one should always measure the actual runtime of an algorithm to obtain real performance results.
Finally, note that the maximal CPU throughput of about 10~GFlops/s shown in \figref{fig:perf_ariel} means about 2.6~Flops/cycle on the 3.8~GHz CPU.
This is quite impressive, but there is still room for improvements as the SIMD registers can generally do up to 4~Flops/cycle, or even more for specific sequences of additions/multiplications.
However, this would involve hand-tuned code which is beyond the scope of this article.

\subsubsection{Results on AMD Opteron}

\begin{figure}[t]
  \subcaptionbox{Performance comparison.\label{fig:perf_trillian}}[0.48\textwidth]{
    \includegraphics[width=0.48\textwidth]{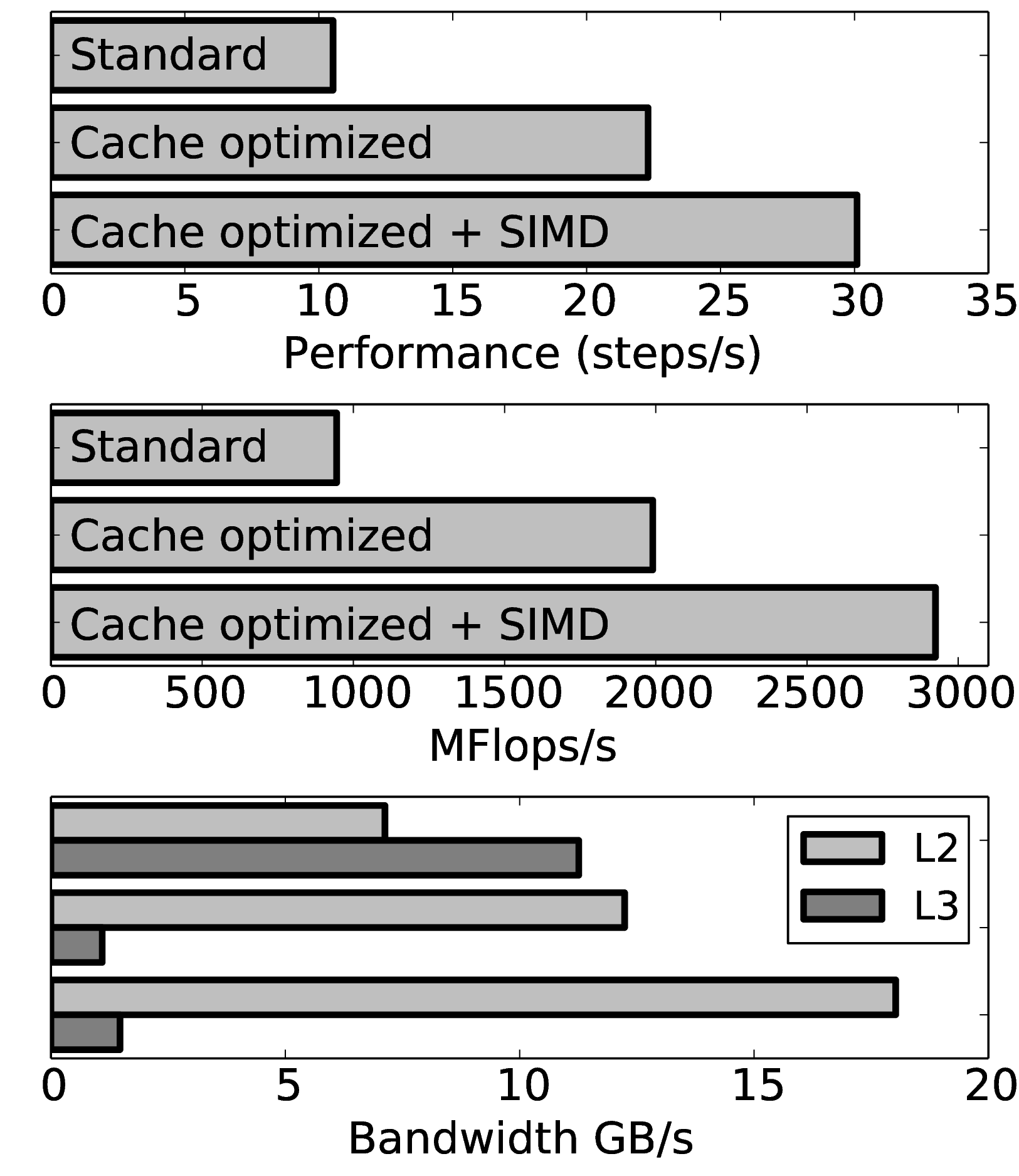}} \hfill
  \subcaptionbox{Granularity of cache optimized version.\label{fig:granularity_trillian}}[0.48\textwidth]{
  \includegraphics[width=0.48\textwidth]{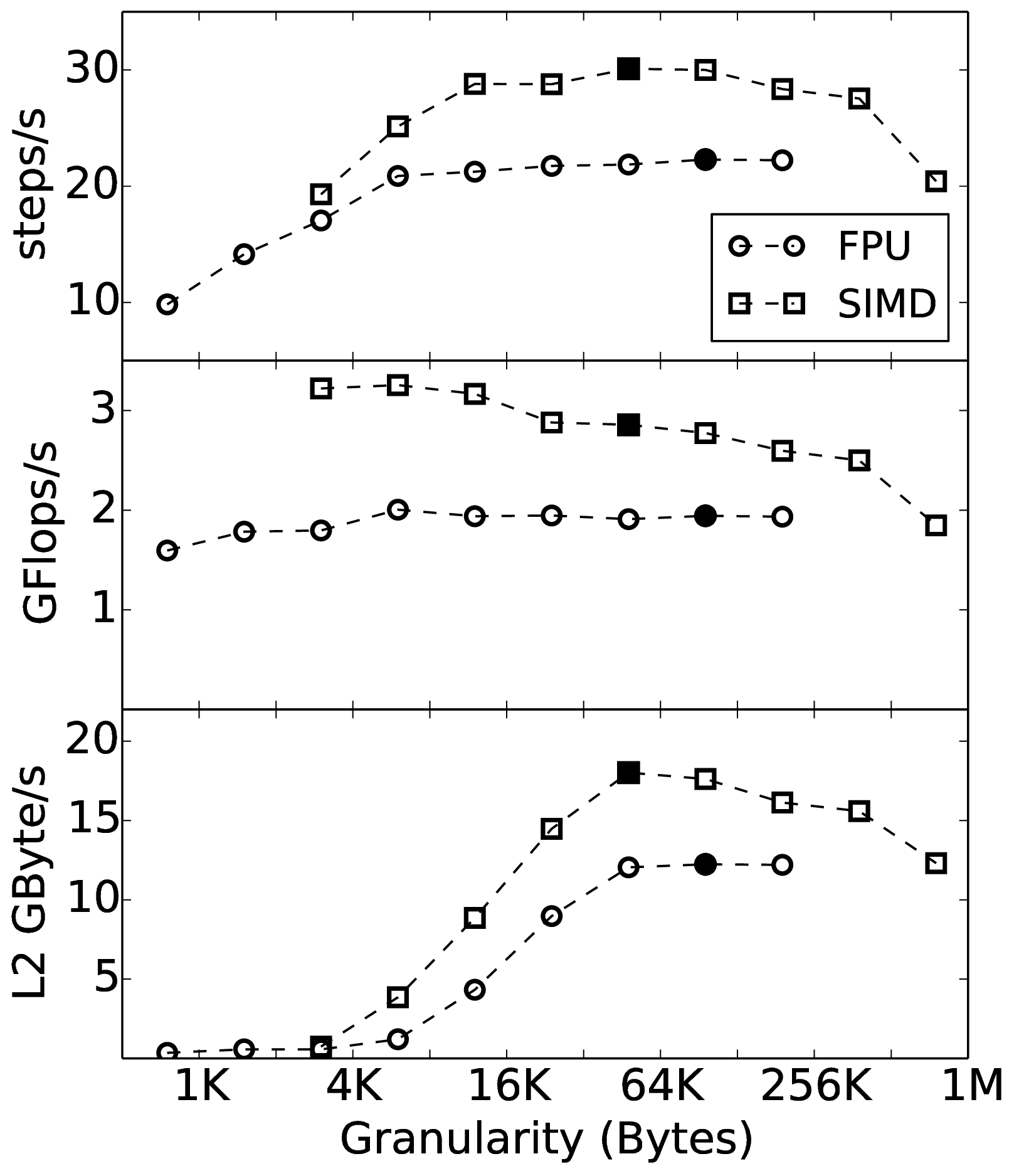}}
  \caption{Performance results for the same simulation as in \figref{fig:perf_ariel_both} on an AMD Opteron 2.1~GHz.\label{fig:perf_trillian_both}}
\end{figure}

We repeated the same tests on an AMD Opteron 6272 with 2.1 GHz and L1/L2/L3 cache sizes of 48KB/1MB/8MB~\cite{conway2010cache}, where we used clang-3.4 to compile the \CXX implementations.\footnote{Test with the g++-4.9 compiler showed similar, but slightly lower ($\sim10$\%) performance results.}
\figref{fig:perf_ariel} shows the performance results for the three different implementations.
Again, the standard implementation suffers greatly from bandwidth limitations with a rather poor performance of only 10 steps per second.
Introducing granularity again greatly improves the performance by reducing the required L3 cache bandwidth.
\figref{fig:granularity_trillian} shows the granularity dependence of the performance for the AMD processor.
For the non-SIMD version we find similar results as for the Intel Xeon before, with a performance increase for increasing granularity at a rather constant rate of CPU throughput of about 2000 MFlops/s corresponding to roughly 1 Flop/cycle.
Hence again we are able to turn the algorithm from cache bandwidth bound to Flops bound by introducing granularity.
Additional utilization of SIMD instructions leads to further performance gains of about 40\%.
The maximal CPU throughput of about 3000 MFlop/s corresponds to roughly 1.4 Flops/cycle, clearly below the 4 Flops/cycle that is possible with this processor's AVX extensions. 
As for the Intel Xeon, this indicates that the simulation does not yet make full use of the available SIMD registers
Again, it might be that with further hand-tuning of SIMD instructions additional performance gains can be reached, but this is beyond the scope of this article.

Nevertheless, we identified the optimal granularity and found a speed up of a factor of three compared to the standard implementation.
The overall speed up is similar to the Intel Xeon, with similar contributions from cache optimization and SIMD usage.
However, the final performance of the Intel Xeon is about another three times better than for the AMD Opteron.
This is partly due to the higher clock speed of 3.8~GHz vs only 2.1~GHz for the Opteron, but also because of the better SIMD utilization for the Intel Xeon.
Indeed, the AMD Opteron only reaches about 1.4~Flops/cycle, roughly half of what was observed for the Intel Xeon above.
Note however, that these results should not be taken as a comparison of the two processors, as they have very different properties, e.g.\ 16 cores in the AMD Opteron vs eight cores in the Intel Xeon.

\subsubsection{Dependence on System Size}

\begin{figure}[t]
  \centering
    \includegraphics[width=0.6\textwidth]{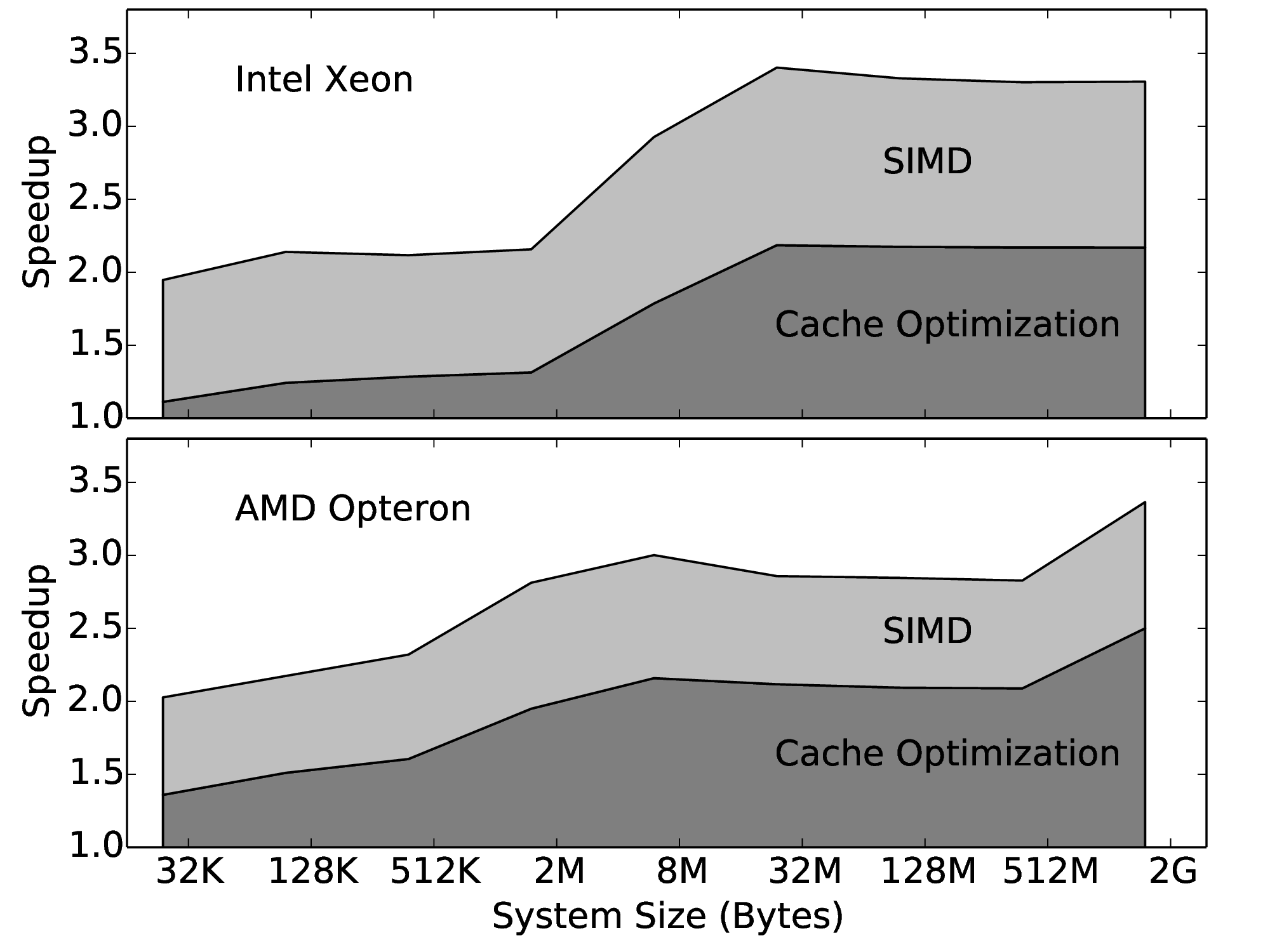} \hfill
  \caption{Speedup gained from cache optimization and SIMD usage compared to standard implementations in dependence of the problem size. The graph shows the performance gain in terms of total run-time of simulating a chain with nearest-neighbor couplings of different size on a Intel Xeon (3.8 GHz, top panel) and AMD Opteron (2.1 GHz, bottom panel). The graphs show the speedup over a standard implementation gained from granularity optimization (dark gray) and additional SIMD instructions (light gray).}
   \label{fig:speedup}
\end{figure}

Finally, we measured the speed up due to cache optimization and SIMD instructions for different chain lengths~${N=2^{10}\dots2^{26}}$.
The result is shown in \figref{fig:speedup} for both the Intel Xeon (upper panel) and the AMD Opteron (lower panel).
The darker regions correspond to the performance gain from cache optimization, while the lighter regions indicate the additional gain from SIMD instructions.
The system size is given in Bytes required to store one instance of the state $\{\vec r_i\}$ by computing ${N\cdot D\cdot 8\,\text{Bytes}}$.
Note that the optimal granularity was observed to be independent of the system size (results not presented), as one expects because it only depends on the available L1/L2 cache size.

For both processors, using the SIMD instructions gives a constant gain independent on system size.
For the Intel Xeon, the influence of cache optimization has a clear increase when the systems get larger than 2~MBytes.
This can be explained by the L3 cache size.
To perform a whole Runge-Kutta-4 step, the algorithm needs to store the state $\{\vec r_i\}$, an intermediate state $\{\vec r'_i\}$ and $s$ rhs evaluations $\{\vec k^j_i\}$.
This gives a total memory requirement of $s+2=6$ states.
That means for the Intel Xeon with 20~MBytes L3 cache, for system sizes larger than about 3~MBytes the L3 chache is not sufficient to contain all required data and the even slower main memory has to be used.
This leads to a further performance decrease of the standard implementation and hence bigger speed up from the cache optimizations.
For the AMD Opteron with an L3 cache size of 8~MBytes, the transition to main memory should happen at system sizes of about 1~MByte, which is also consistent with \figref{fig:speedup}.
Furthermore, there we observe another increase of the speed up at system sizes reaching about 2~GBytes.
This is probably due to the fact that the main memory of this machine is divided into NuMA domains (Nonuniform Memory Access).
So for very large systems, the single-core simulation requires memory outside of its NuMA domain, which induces further bandwidth limitations and hence even bigger speed up due to cache optimization.

\section{Implementation}

All simulations were implemented in \CXX and the sources are made publicly available~\cite{github_source}.
For the numerical Runge-Kutta-4 algorithm the Boost.odeint library was used~\cite{Ahnert_Mulansky_odeint:11,odeint_url}.
It provides modern, generic implementations of ODE solvers and greatly simplified the implementation of the cache optimized algorithm.
For example, Boost.odeint performs the required memory allocations for the temporary results of the RK4 steps and also supports an easy way to introduce granularity with the help of a range interface.
To efficiently employ the SIMD registers, $\text{NT}^2$'s SIMD library~\cite{esterie2014boost} (proposed Boost.SIMD) was used.
It provides a \CXX abstraction to the low-level SIMD registers and allows for a very fast and easy porting of the simulation to SIMD.
By using this library, the source code is generic in the sense that it will be decided at compile-time, and not in the source code, which SIMD extensions are going to be employed (e.g.\ SSE3, SSE4 or AVX).
Furthermore, the generic implementation of Boost.odeint allowed us to readily use SIMD data types, and therefore SIMD registers, without the need to re-implement the Runge-Kutta-4 algorithm with SIMD instructions.

For more details of the implementations we refer to the source codes~\cite{github_source}.
We only emphasize here that the introduction of granularity as well as the SIMD usage can be done with reasonable programming effort when using Boost.odeint and Boost.SIMD.
For example, the simulation of the coupled Roessler system with granularity and SIMD instructions requires merely about 220 lines of \CXX code, only about 100 lines more than a basic standard implementation with Boost.odeint.
Hence, applying the techniques presented here in ODE simulations with \CXX is certainly achievable in most situations.

\section{Conclusions and Outlook}

In this article, we have presented a technique to improve the performance of large-scale ODE simulations with nearest-neighbor coupling based on explicit Runge-Kutta schemes.
The main performance bottleneck for such simulations is the cache and memory bandwidth which prevents the full employment of CPU throughput due to the lack of data available to the CPU.
We are able to overcome this problem by introducing a granularity to the algorithm which leads to a more cache efficient implementation.
To ensure the correctness of the algorithm, we introduced overlap computations, which effectively increased the required computational effort.
However, performance tests on two processors, an Intel Xeon and an AMD Opteron, showed that for optimal granularity, this is clearly out-weighted by the better cache utilization.
With this improvement, we again arrived at an implementation that is bound by the CPU throughput instead of cache/memory bandwidth.
We were then able to further increase the performance by employing SIMD instructions.
For both processors we found a total performance increase of up to a factor of three, depending on the system size (cf.\ \figref{fig:speedup}).
We conclude that for large-scale simulations, the techniques presented here should be considered as a valuable option to increase simulation performance, typically even before exploring multi-core parallelization.

Note that from Figures~\ref{fig:granularity_ariel} and \ref{fig:granularity_trillian} one finds that the optimal granularity roughly corresponds to the L1 cache sizes of the used processor.
Hence, it is possible to automatically choose the optimal granularity without employing a performance study by adjusting the granularity to the L1 cache size of the target machine.

However, the cache optimizations are only promising if the program indeed suffers from memory/cache bandwidth limitations.
In cases, for example, where the rhs of the ODE involves more complicated expressions that require several CPU cycles to be computed, the cache/memory bandwidths might not impose a bottleneck and hence introducing granularity would not lead to any performance gain.
In such situations, one should directly try to utilize SIMD instructions to increase the CPU throughput and obtain better performance this way.

To ensure the correct computation in cases with granularity we introduced overlap computations.
The presented strategy only works for nearest-neighbor coupling.
In general, however, the same idea could also be applied in more complicated situations.
In case of next-nearest-neighbor couplings, for example, simply twice as long overlaps have to be introduced, and similarly with longer coupling ranges.
In principle, also for situations with distant, but sparse couplings granularity can be implemented.
But this is more complicated as one has to keep exact track of the error propagation in the Runge-Kutta scheme.
Furthermore, if the coupling is not sparse enough, the gain from cache optimization might not outweight the additional overlap computations anymore.
Nevertheless, with a factor three speed up, the cache optimization techniques presented here proved to be highly valuable for nearest-neighbor coupling and can potentially be generalized to more complex situations as well.

\section{Acknowledgements}
I thank Karsten Ahnert and Denis Demidov for fruitful discussions.
Moreover, I thank Denis Demidov for proof reading the manuscript.
Finally, I thank Hartmut Kaiser and the \stellar group for hospitality and computer \underline{}cluster access.

This work was supported by the European Comission through the Marie Curie Initial Training Network \emph{Neural Engineering Transformative Technologies} (NETT) under the project number 289146.

\bibliographystyle{siam}
\bibliography{olsos}

\end{document}